\documentclass[a4paper,10pt]{article}

\usepackage[noadjust]{cite}
\usepackage{url}
 
\usepackage{tex_andy}
\usepackage{algorithms_andy}
\usepackage{equations_andy}
\usepackage{theorems_andy}
\usepackage{figures_andy}
\usepackage{sections_andy}
\usepackage{tables_andy}
\usepackage{graphicx,eepic,epic,epsfig}
\usepackage{times}
\usepackage{andreas}

\usepackage{tabularx}
 
\usepackage{k_tex,misc_k}

\setcommand{\Intvl}[2]{\:[#1, #2)\:}
\makeatletter
\@ndyTheorem{Invar}{Invariance}{Invariance}     
\makeatother
 
\title{A Static Data Structure for Discrete Advance Bandwidth Reservations on the Internet}
\author{
    Andrej Brodnik 
  \thanks{
    Department of Theoretical Computer Science,
    Institute of Mathematics, Physics, and Mechanics, Ljubljana, Slovenia
  }
$\ ^\dag$
  \and
    Andreas Nilsson
  \thanks{
    Department of Computer Science and Electrical Engineering,
    Lule{\aa} University of Technology, Lule{\aa}, Sweden
  }
}
\date{}

\begin{document}
\maketitle
\begin{abstract}
  In this paper we present a discrete data structure for 
  reservations of limited resources. A reservation is defined as
  a tuple consisting of the time interval of when the resource should
  be reserved, $I_R$, and the amount of the resource that is reserved, 
  $B_R$, formally $R=\{I_R,B_R\}$. \\  
  The data structure is similar to a segment tree. The maximum spanning 
  interval of the data structure is fixed and defined in advance. The 
  granularity and thereby the size of the intervals of the leaves is 
  also defined in advance. The data structure is built only once. 
  Neither nodes nor leaves are ever inserted, deleted or moved. 
  Hence, the running time of the operations does not depend on the 
  number of reservations previously made. The running time does not
  depend on the size of the interval of the reservation either. 
  Let $n$ be the number of leaves in the data structure.
  In the worst case, the number of touched (i.e. traversed) nodes is in 
  any operation $O(\log n)$, hence the running time of any operation is also
  $O(\log n)$
\end{abstract}
\section{Introduction}\sectlabel{intro}
The original, never published version of this paper was called 
\emph{``An Efficient Data Structure for Advance Bandwidth Reservations 
on the Internet''}. The original paper was referred to in the paper 
``Performance of QoS Agents for Provisioning Network Resources'' 
(\cite{ schelen99}) by Schel\'{e}n et. al. under the reference 
number 14, but the reference should really be changed to the
current paper.

\section{Definition of the problem}\sectlabel{ProblemDefinition}
The problem we deal with, we call \emph{``The
Bandwidth Reservation Problem''}. A reservation is a time interval during
which we reserve constant bandwidth. The solution given here works 
in a discrete bounded universe.

By a discrete bounded universe we mean a universe with a limited duration and
a fixed time granularity. By using a fixed granularity we have
divided the time into time slots (frames). We use a slotted time in hope to
get a smaller data structure, faster operations and hence gain benefits of a
larger aggregation. This hope is inspired by the fact that problems are 
generally easier to solve in a bounded (discrete) universe than in the 
general case (\cite{Brodnik-thesis}).

We observe that the bandwidth reservation problem is constrained by the physical
world and therefore no reservations will occur in the past and very few
in a distant future. 

Throughout the paper we use the following notation:

\begin{itemize}
  \item 
    We have a bounded maximum interval $M$ starting at $S_M$ and ending
    at $E_M$, hence $M=[S_M,E_M]$. $M$ is divided into fixed size time 
    slots of size $g$. The size of the interval $M$ is denoted by $\left| M\right| $.

  \item
    In general, a discrete interval, $I$, is defined as the duration between a starting
    point $S$ and an ending point $E$, and the interval is divided into 
    discrete slots of size $g$. In short, $I=[S,E]$. Moreover, since 
    $S_M\leq S<E\leq E_M$ $I\subseteq M$.
    
  \item
    The bandwidth is denoted by $B$. A reservation, $R$, is defined by an 
    interval $I$ and a (constant) amount of reserved bandwidth $B$, 
    during $I$. In short, reservation is a tuple $R=\{B,I\}$. 
    Items related to reservations are denoted by a subscript, e.g. 
    $R=\{B_R,I_R\}$.
    
  \item
    A data structure storing reservations made is denoted by \calD.  An 
    item related to a ``query'' toward the data structure is denoted 
    by a subscript $Q$, e.g. $I_Q=[S_Q,E_Q]$.

\end{itemize}

The bandwidth reservation problem defines three
operations. First, we have a query operation
that only makes queries of the kind: ``How much bandwidth is reserved at most 
between time $S$ and time $E$?''. Further, we have update operations: an 
insertion of a new reservation and a deletion of a reservation
already made. Formally these operations define:

  \bDefn{definition1} 
    We have a bounded maximum interval $M$ divided into time slots 
    of size $g$. Let a reservation, $R=\{B_R,I_R\}$, be on an
    interval $I_R\subseteq M$ with an associated bandwidth, $B_R$. 
    Then the \emph{bandwidth reservation problem} requires the following operations:

    \begin{description}
      \item  
        $\mathtt{Insert}(\calD\ ,R)$, which increases the reserved bandwidth
        during the interval $I_R$ for $B_R$.

      \item  
        $\mathtt{Delete}(\calD,R)$, which decreases the reserved bandwidth
        during the interval $I_R$ for $B_R$.
      \item
        $\mathtt{MaxReserved}(\calD,I_Q)$ which returns the maximum reserved 
        bandwidth, during the interval $I_Q$.

    \end{description}

  \eDefn

  Note, deletion is the same as an insertion but with a negative bandwidth.

\subsection{Background of the problem}

  The bandwidth reservation problem is not so well studied in the literature.
  On the other hand, two related problems, the partial sum problem 
  (\cite{Fredman82}, brief in \cite{HusRau-ICALP-98}), and the prefix sum problem 
  (\cite{Fredman82}), are. In the partial sum problem we have an array 
  $V(i),1\leq i\leq n $ and want to perform these two operations: $(1)$ $\mathtt{update}$: 
  $V(i)=V(i)+x$; and $(2)$ $\mathtt{retrieve}$: $\sum_{k=1}^mV(k)$ for arbitrary
  values of $i$, $x$ and $m$. There is only a slight difference between the partial 
  sum problem and the prefix sum problem, in the prefix sum problem
  the query always starts at the beginning and
  in the partial sum problem the queries are for an arbitrary interval.
  
  In our solution we will use a data structure similar to segment trees (\cite{Meh84a}).
  Segment trees  represent a method for storing set of 
  intervals. 
  For instance, we have $m$ intervals with $n$ unique starting or ending points.
  The segment tree is then an efficient data structure for storing those intervals and answering queries 
  over which of the $m$ intervals spans the query interval.
  Formally, let \calS\ denote a set of $m$ intervals with $n$ unique starting and ending points.
  Let $Z_i$ be a starting or ending point for an interval in \calS.
  Let $U$ be our universe where $U=\{Z_1,Z_2,...,Z_n\}$ given that $Z_{i-1}<Z_i$ where $1<i\leq n$.
  The leaves in the segment tree correspond to the intervals:
  $\left( -\infty ,Z_1\right) ,\left[ Z_1,Z_1\right] ;
   \left( Z_1,Z_2\right) ,\left[ Z_2,Z_2\right] ;...;            \linebreak
   \left( Z_n,+\infty \right) $ as shown in \figref{figures0}.
  \\
  \figinclgraph{figures0}{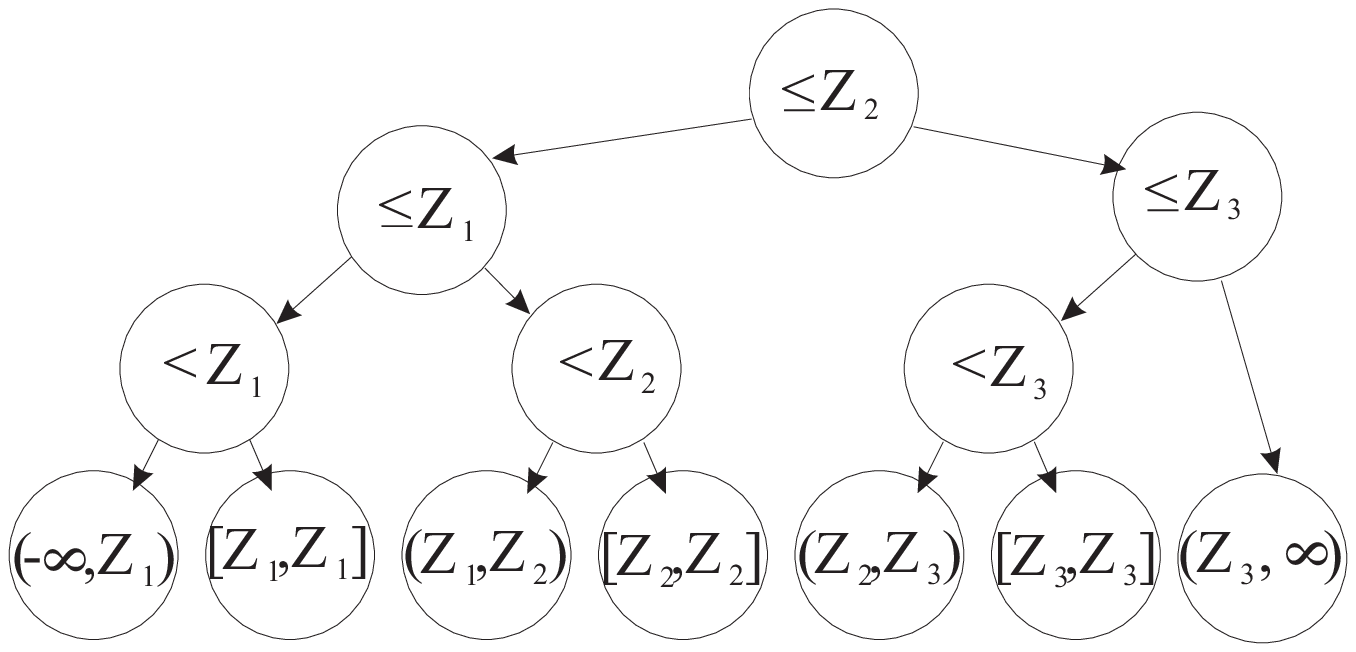}{width=\linewidth}
  {An example of a segment tree.}
  \label{figures0}
  \\
  An internal node represents the interval of the tree rooted at it and 
  it also stores the information about the interval of its left subtree. 
  The interval of the node is the union of the intervals of its children.
  Each node (and a leaf) contains a list of pointers to previously inserted intervals that 
  completely cover the interval of the node but not the interval of the node's parent.
  During the insertion of the interval $I$, the pointer to the interval $I$ 
  is inserted in a node $N$'s list, if all children of $N$ have their corresponding intervals 
  within $I$ and the parent of $N$ does not. Consequently, pointers to an interval
  are stored at maximum two nodes on each level of the tree.
  The segment tree has $2n+1$ leaves and $2n$ nodes. Since the nodes in the tree have 
  constant number of children the height of the tree is $O\left( \log n\right) $.
  This is also the complexity of an insertion and a query.

\section{Solution}\sectlabel{Discrete_Solution}
\newcommand{\Abs}[1]{\left| #1 \right|}
Our solution is a modified segment tree (\cite{Meh84a}). 
In our data structure, Advanced Segment Tree (AST), each node represents one time interval.
Every node in the tree consists of the interval it represents,
pointers to each of the node's children, and two values 
(described more thoroughly further down).

The interval to which the root corresponds is $M$.
Let $L$ denote the number of levels that the data structure consists of.
All nodes on level $l$ have time intervals of the same size and they do not intersect.
They follow consecutively one another.
This means that the union of all intervals on level $l$ is $M$.
Each level has a divisor
that tells the number of children that a node on that 
particular level has. 
The divisors are gathered up from the root to the 
leaves in a set $X=\{X_1,X_2,...,X_{L-1}\}$, where $X_1$ 
is the number of the root's children.
The divisor $X_l$ does not only tell the number of children that a
node has, but also the size of the interval $\Abs{M_l}$ on 
level $l$:
\bEqu{Intervalsize}
 \Abs{M_l} = \left\{ \begin{array}{ll} \Abs M & i=1 \\
      \frac{\Abs{M_{l-1}}}{X_l} & 1<l\leq L \end{array}
  \right.
\eEqu
Consequently the number of nodes on level $l$ is
\bEqu{n}
  n_l=\left\{ \begin{array}{ll} 1 & l=1 \\
      \prod_{i=1}^{l-1}X_i & 1<l\leq L  \enspace . \end{array} 
  \right.
\eEqu
and the number of leaves of the complete data structure is
\bEqu{n}
  n=n_L=\prod_{i=1}^{L-1}X_i  \enspace .
\eEqu


The divisors $X_i$ must be set so that $g=\frac{\Abs{M}}{n}$, where $n$ is defined in \rEqu{n} and 
where $g$ is the time granularity (size of the leaves). 
Hence, the choices of $\Abs{M}$, $g$, $L$, $n$ and $X$ are related.
For instance, choosing $\Abs{M}$ to be a prime number
makes the tree consist of only two levels, 
the top and the leaf level. We get the simplest tree when $\Abs{M}=2^L\cdot g$,
i.e. 
$X=\{X_l\mid X_l=2$, \textrm{for} $0<l<L\}$.
Note that the fundamentals of the data 
structure do not change if the values of the divisors change. There 
will however be a deterioration in performance for each added level.
The tree is only built once and therefore the tree is always perfectly
balanced. 
There is a difference between the segment tree and our data structure regarding the leaves. 
In the segment tree there is a leaf for the open interval, $(X_{i-1},X_i)$, as well as the 
closed interval, $\left[ X_i,X_i\right] $, but in our data structure leaves represent semi-open 
intervals $\left( X_i,X_{i+1}\right] $. To describe our data 
structure we use the following notation:

\begin{itemize}
  \item 
    Let $N$ denote ``the current node'' during a description of a traversal of the tree. 
    Let $N_L$ denote the leftmost child of $N$, and $N_R$ the rightmost child.

  \item 
    Each node, $N$, stores the interval $I_N=[S_N,E_N]$ that the node subtends.
    
  \item  
    Each node, $N$, stores the amount of bandwidth, $nv_N$, that was reserved 
    over exactly the whole interval $I_N$.
  \item

    
    Each node $N$ also stores the maximum value of reserved bandwidth excluding the value $nv_N$ 
    on the interval $I_N$. This maximum value is denoted as $mv_N$.
  
\end{itemize}

\bAlg{fig:ATIM-code-definitions}{pascal}{Advance Tree Definitions in C}
\#define Split \{2,2,2,2,2,3,2,2,2,3,2,2\}
\#define MaxSplit 3
typedef struct NODE \{
   interval\_type Interval;
   int           node\_value;
   int           max\_value;
   struct NODE *Children[MaxSplit];
\} AST\_type;

\eAlg

In \figref{figures1} is given an example of how to make the data structure 
subtending a 32-day month, with $g$ representing $5$ minutes.
All nodes have $2$ or $3$ children in order to have the wanted interval sizes. 
Hence, in \figref{figures1} $X=\{2,2,2,2,2,3,2,2,2,3,2,2\}$.

\figref{figures2} presents an example of a bandwidth utility graph. The 
graph shows the amount of bandwidth that is reserved over time. In \figref{figures3}
we are showing the values in the tree corresponding to the example graph from \figref{figures2}.
\figref{figures3}, also shows how $mv$ is calculated.

  \figinclgraph{figures1}{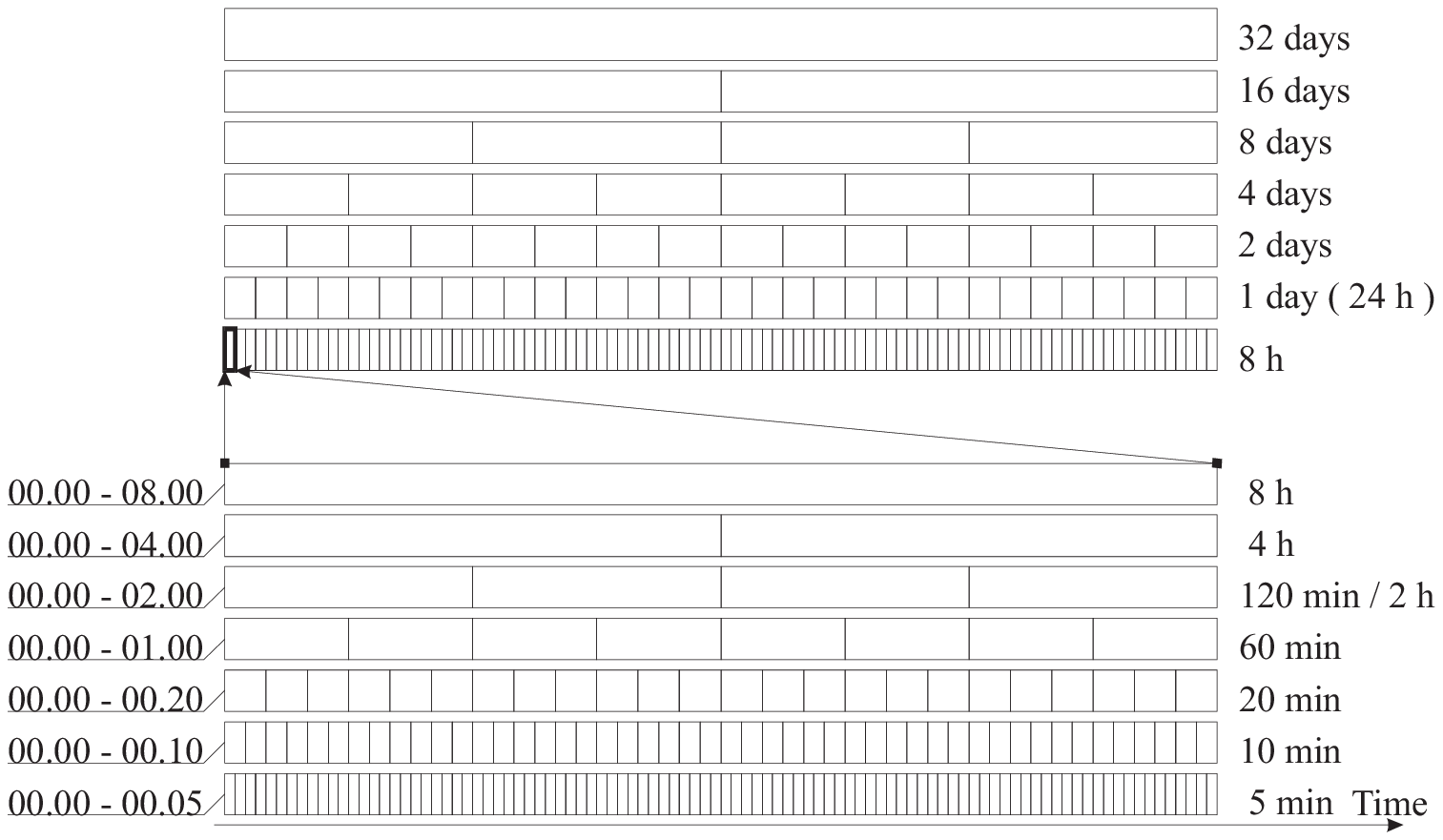}{width=\linewidth}
  {An example of the data structure.}
  \label{figures1}

  \figinclgraph{figures2}{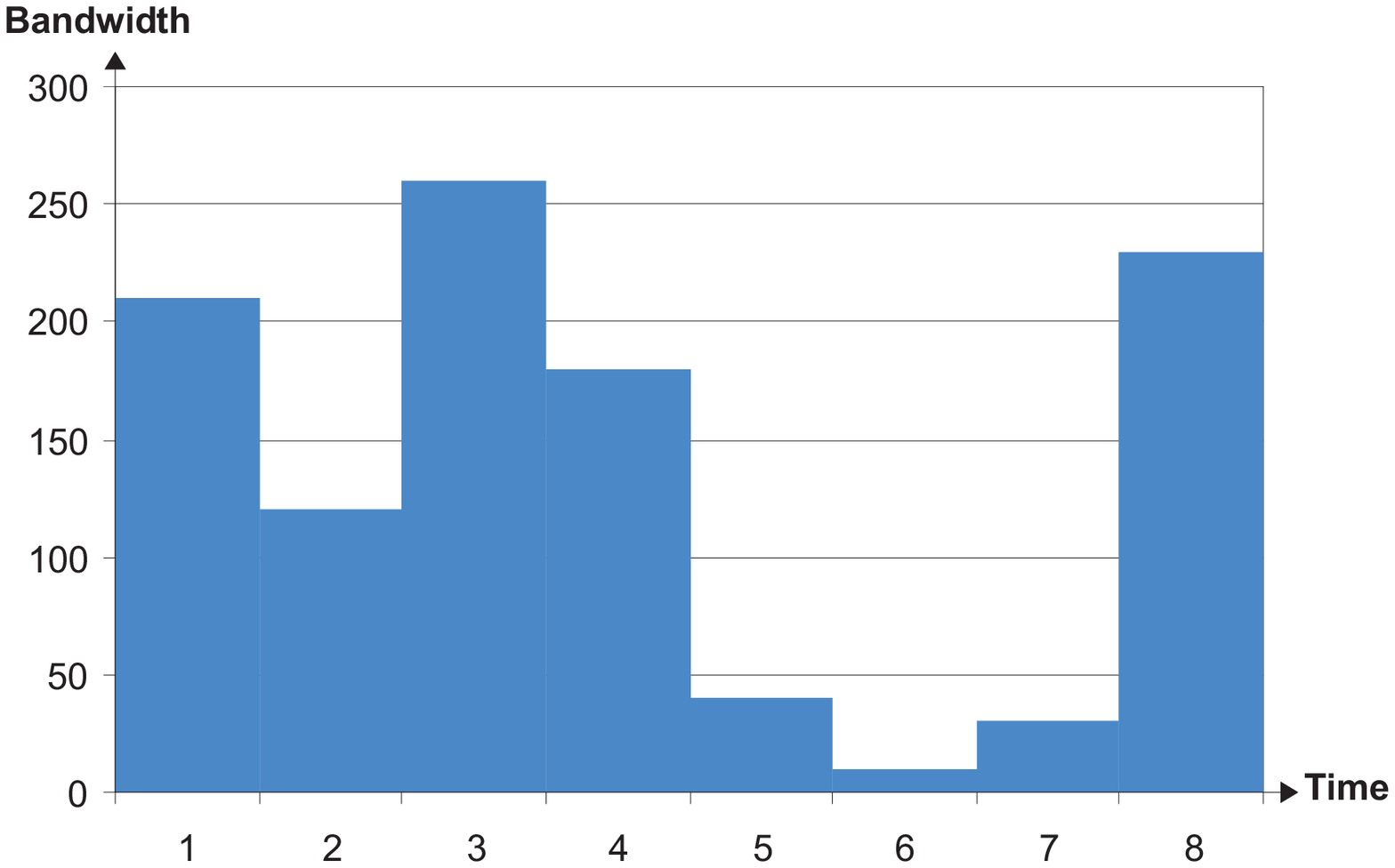}{width=\linewidth}
  {Bandwidth - Time graph.}
  \label{figures2}

  \figinclgraph{figures3}{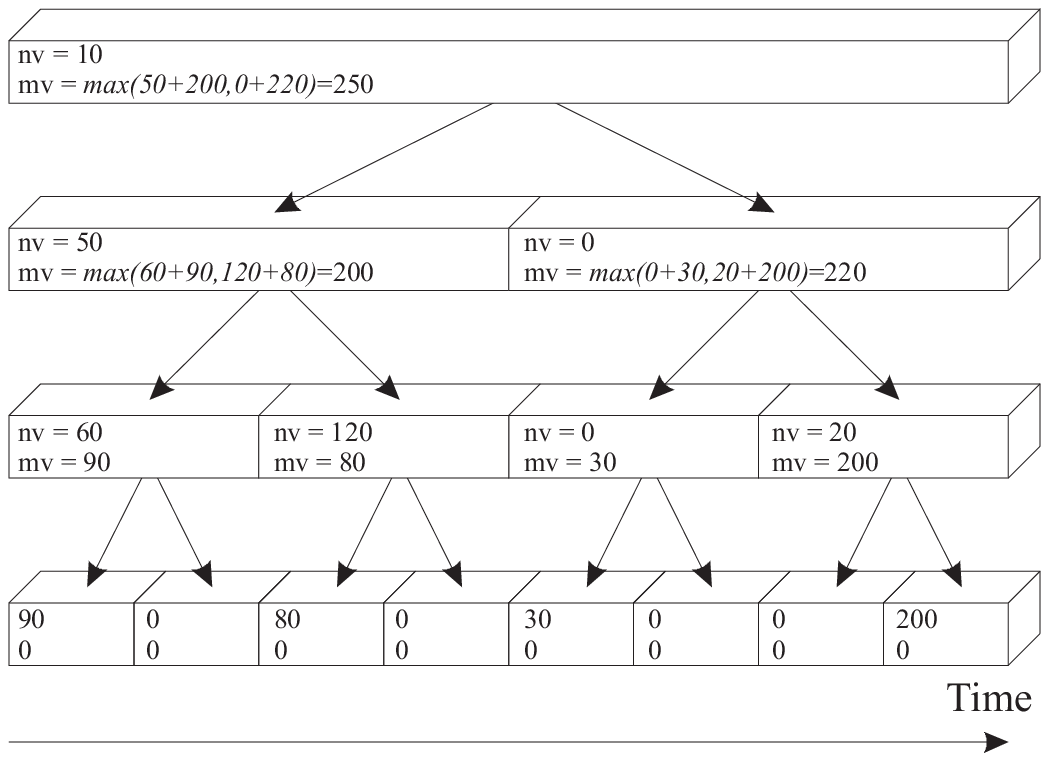}{width=\linewidth}
  {An example showing values in the data structure.}
  \label{figures3}
  
To describe the operations from \rDefn{definition1} we
use the data structure formally defined in \rAlg{fig:ATIM-code-definitions}.
We start by describing the function \linebreak
$\mathtt{MaxReserved}(N,I_Q)$.
\begin{itemize}
  \item
    If the interval of node $N$ satisfies $I_Q=I_N$ (i.e. $S_N=S_Q$ and $E_N=E_Q$) then $nv_N + mv_N$
    is returned as the result.

  \item
    If $I_Q$ is entirely contained within the interval of child $N_C$ of $N$, then the returned value will be:
    $nv_N+$ $\mathtt{MaxReserved}(N_C,I_Q)$.

  \item
    If $I_Q$ spans the intervals of more than one child of $N$, $I_Q$ is divided 
    into one part for each of the $m$ children of $N$ which intervals $I_Q$ at least partially spans - i.e.\
    $I_{Q_{i}}=I_Q\cap I_{N_{Ci}}$, \textrm{for} $1\leq i\leq m$ 
    where $I_{Q_1}$ 
    is the leftmost interval and $I_{N_{C1}}$ is the interval of the leftmost 
    child of $N$ that has an interval that at least partially spans $I_Q$. 
    The \figref{figures4} illustrates the split of $I_Q$ into smaller intervals.


\figinclgraph{figures4}{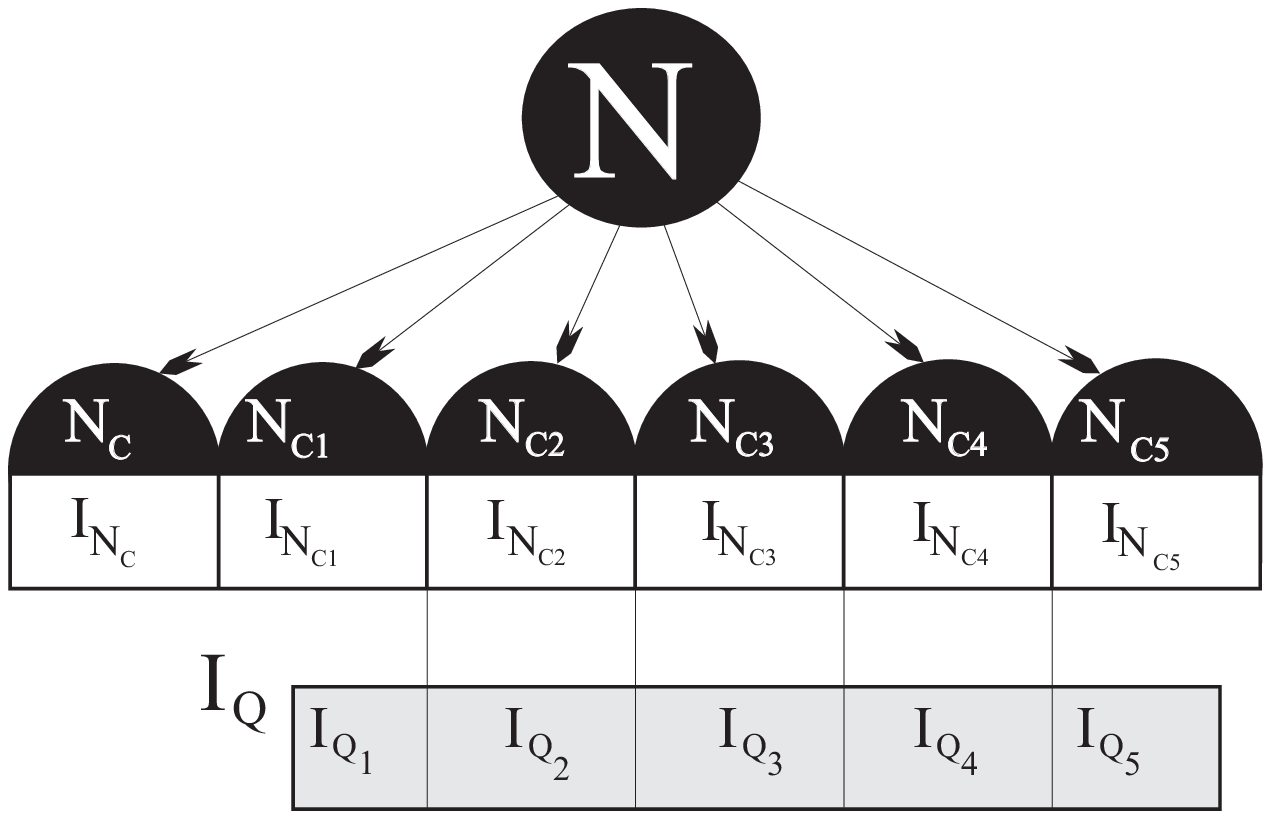}{width=\linewidth}
{An example showing a split of an interval (circles represent nodes and rectangles represent intervals).}
\label{figures4}

    The returned value will be
    \[ \max\limits_{1\leq i\leq m}\left( \mathtt{MaxReserved}\left( N_{Ci},I_{Qi}\right) \right) \]
    and it can be more efficiently computed as:
\bEqu{MaxReserved}    
  \begin{array}{rll} 
   \max\left(  \right. &  \mathtt{MaxReserved}\left( N_{C1},I_{Q1}\right)  &  ,      \\
               &  \max\limits_{1<i<m}\left( nv_{N_{Ci}}+mv_{N_{Ci}}\right)  &  ,      \\
               &  \mathtt{MaxReserved}\left( N_{Cm},I_{Qm}\right)           &  \left. \right)  \end{array}
\eEqu

\end{itemize}

In \figref{figures5} an insertion of a reservation of $B_R$ units of bandwidth is shown.
The figure also shows which nodes will be touched when a 
$\mathtt{MaxReserved}$ function is invoked with the same query interval. The $\mathtt{MaxReserved}$ 
function is formally defined in \rAlg{fig:ATIM-code-MaxReserved}.


\bAlg{fig:ATIM-code-MaxReserved}{pascal}{Advance Tree MaxReserved in C}
bw\_type \_MaxReserved(AST\_type *node,int start, int stop, 
        int iLevel) \{
   bw\_type    MaxBelowNode, MaxBelowNode;
   int        i1;
   AST\_type *pRunningNode;
      
   if (start==node->Interval.Start \&\& stop==node->Interval.End) 
      return (node->node\_value + node->max\_value);
   else \{
      MaxBelowNode = 0;
      for (i1=0;i1<Split[iLevel];i1++) \{
         pRunningNode = node->Children + i1;
         if ( start < pRunningNode->Inteval.End ) \{
            if (pRunningNode->Interval.End < stop) \{
               MaxBelowChild = \_MaxReserved(pRunningNode, start, 
                       pRunningNode->Interval.End, iLevel+1);
               if ( MaxBelowChild > MaxBelowNode )
                  MaxBelowNode = MaxBelowChild;
               start = pRunningNode->Interval.End;
            \} else \{
               MaxBelowChild = \_MaxReserved(pRunningNode,start,
                                            stop,iLevel+1);
               if ( MaxBelowChild > MaxBelowNode )
                  MaxBelowNode = MaxBelowChild;
               break;
            \}
         \} /* if */
      \} /* for */
      return( node->node\_value + MaxBelowNode );
   \}
\}

bw\_type MaxReserved(AST\_type Data, interval\_type I) \{
   \_MaxReserved(\&Data,I.Start,I.End,0);
\}

\eAlg

 The function $\mathtt{Insert}(N,\{I_R,B_R\})$ works in a similar way as the  \linebreak
 $\mathtt{MaxReserved}$ 
 function. $\mathtt{Insert}$ must also verify that the inserted reservation 
 does not result in an over-reservation of bandwidth. This verification can be 
 done by making the reservation, then performing a $\mathtt{MaxReserved}(N,I_R)$ query, 
 and finally comparing with the maximum reservable bandwidth on the link. If an 
 over-reservation occurs the reservation must be removed. More 
 efficient is to 
 let the $\mathtt{Insert}$ function perform the check during its execution. We will describe 
 the recursive $\mathtt{Insert}$ function without integration of $\mathtt{MaxReserved}$ functionality, 
 which inclusion is trivial and is shown in \Sectref{section:Early rejection}.
 
\begin{itemize}
  \item
    If the interval of node $N$ satisfies that $I_R=I_N$ (i.e. $S_N=S_R$ and $E_N=E_R$), then $nv_N$ 
    is increased by $B_R$. 

  \item
    If $I_R$ is entirely contained within the interval of one child $N_C$ of $N$, then 
    the function $\mathtt{Insert}(N_C,\{I_R,B_R\})$ is called and when it returns the $mv_N$ 
    is updated according to the equation 
\bEqu{Insert}    
    mv_N=\max\limits_{1\leq i\leq k}\left( nv_{Ci}+mv_{Ci} \right) \enspace ,
\eEqu
    where $k$ is the number of children that $N$ has.

  \item
    If $I_R$ spans the intervals of more than one child of $N$, $I_R$ is divided exactly as the query interval
    into one part for each of the $m$ children of $N$ which intervals $I_R$ at least partially spans
    - i.e. $I_{R_{i}}=I_R\cap I_{N_{Ci}}$, \textrm{for} $1\leq i\leq m$. 
    The $\mathtt{Insert}$ function is called once for each of the $m$ children $N_{Ci}$,
    $\mathtt{Insert}(N_{Ci},\{I_{R_i},B_R\}), 1\leq i\leq m$.
    When the calls return, the $mv_N$ is updated as shown in \rEqu{Insert}.
    
\end{itemize}
In \figref{figures5} an insertion of a reservation and the calculation of the new $nv$'s 
and $mv$'s are shown. The $\mathtt{Insert}$ function is 
formally defined in \rAlg{fig:ATIMcodeInsertion}.

\bAlg{fig:ATIMcodeInsertion}{pascal}{Advance Tree Insertion in C}
void \_Insert(AST\_type *node,int start, int stop,int bandwidth, 
              int iLevel) \{
   int        ml, mr, i1;
   AST\_type *pRunningNode;
      
   if (start==node->Interval.Start \&\& stop==node->Interval.End) 
      node->node\_value = node->node\_value + bandwidth;
   else \{
      for (i1=0;i1<Split[iLevel];i1++) \{
         pRunningNode = node->Children + i1;
         if ( start < pRunningNode->Inteval.End ) \{
            if (pRunningNode->Interval.End < stop) \{
               \_Insert(pRunningNode, start, 
                       pRunningNode->Interval.End, bandwidth,
                       iLevel+1);
               start = pRunningNode->Interval.End;
            \} else \{
               \_Insert(pRunningNode,start,stop,bandwidth,
                        iLevel+1);
               break;
            \}
         \} /* if */
      \} /* for */        
      ml = node->max\_value;
      for (;i1>=0;i1--) \{
         mr = node->Children[i1].node\_value+
              node->Children[i1].max\_value;
         if ( mr>ml )
            ml=mr;
      \}
      node->max\_value=ml;
   \}
\}
void Insert(AST\_type Data, reservation\_type R) \{
   \_Insert(\&Data,R.Interval.Start,R.Interval.End,R.BW,0);
\}

\eAlg

The $\mathtt{Delete}(N,\{I_R,B_R\})$ is implemented as a call of $\mathtt{Insert}$ with $B_R$ \linebreak
negated, \rAlg{fig:ATIM-code-delete}.

\bAlg{fig:ATIM-code-delete}{pascal}{Advance Tree Delete in C}
void Delete(AST\_type Data, reservation\_type R) \{
   \_Insert(\&Data,R.Interval.Start,R.Interval.End,0-R.BW,0,0);
\}

\eAlg

The $\mathtt{Insert}$ function as well as the $\mathtt{MaxReserved}$ function 
only traverses the tree twice from 
the top to the bottom.
Once for the rightmost part of the interval $I_Q$ and once for the
leftmost part. For the middle part of the interval the recursion 
never goes deeper than $1$ level. The update of the $mv$ values is done during 
the traversal so no further work is needed.
For a tree where every node only has two children,
both functions will touch at the most $4\cdot (\lg n)-7$ 
nodes. For a tree with other divisors the constants are different but still the running time remains $O(\log n)$. 
Even if the check for over-reservations is implemented as a 
separate call to the $\mathtt{MaxReserved}$ function the running time 
remains within $O(\log n)$. 
\figinclgraph{figures5}{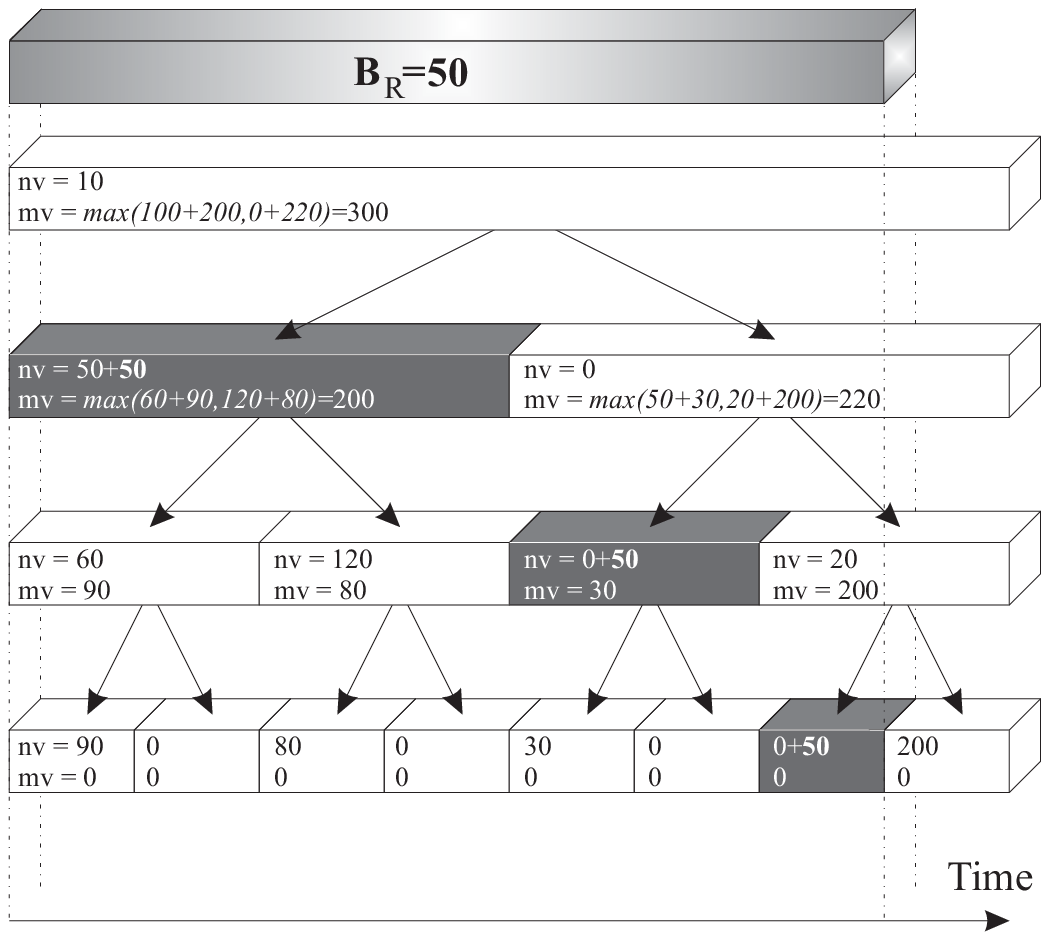}{width=\linewidth}
{An example showing an insertion in the data structure.}
\label{figures5}


\bTheo{AllOperations}
  The running time for all operations solving the bandwidth 
  reservation problem as defined in \rDefn{definition1} and using AST, is $O(\log n)$.
\eTheo

\subsection{Implicit data structure}

    The obvious way to implement our data structure is to use pointers between the 
    nodes. Since our data structure is built only once and the nodes never change, 
    it is possible to store the data structure in an implicit way in an array. In 
    an implicit data structure the positions of the nodes are implicitly defined.
    We use the set $X$, which tells the number of children each node has on level $l$ 
    (see \rEqu{Intervalsize}), to compute the positions of the nodes.
    Once the array is built we can use $X$ to calculate 
    the index of the node instead of using pointers to traverse
    the tree.
    To calculate $\delta _l$, the number of nodes on level $l$ and above, we get from \rEqu{n}:
\bEqu{Arraysize}    
  \delta _l = \left\{ \begin{array}{ll} 1 & l=1 \\
     \sum\limits_{j=2}^l\left( \prod\limits_{i=1}^{j-1}X_i\right) +1 & 1<l\leq L \end{array}
  \right.
\eEqu
    We index the elements in the array from $1$ to $\delta _L$ 
    and order the nodes level by level (cf. standard heap order).
    Consequently, the index of the first element on level $l$, is the number of 
    nodes in the tree on all previous levels, plus $1$, i.e. 
    $\delta_{l-1}+1$. We store these values in vector $\sigma$.
    
    The number of nodes between the node with the index $N$ on level $l$ and 
    the first node on level $l$ is given by $N-\sigma _l$. The number of nodes between 
    the first node on level $l+1$ and $N$'s first child is given by $\left( N-\sigma_l \right) \cdot X_k$.
    The index of the first node on level $l+1$ is given by $\sigma_{l+1}$. The number of children 
    that $N$ has is $X_l$ hence $N$'s children indexes $\gamma$ are:
\bEqu{Indexofchild}    
    \gamma =\sigma_{l+1} +\left( N-\sigma_l \right) \cdot X_l+c, \quad \textrm{for} \quad 0\leq c<X_l
\eEqu
    By using an array instead of using pointers we save the memory for two 
    pointers per node. The execution will be faster due to one memory probe 
    less since vectors $X$ and $\sigma$ will be in cache.

\subsection{Improvements}

  In this section we describe some performance improvements which should be seen as 
  hints to an implementer of our data structure.

  \subsubsection{Choice of the intervals}

    The idea is, to make proper choices about the intervals to improve the running time of the functions.
    The choices to be made are regarding the duration as well as starting times and ending times.
    If we are dealing with man made reservations we observe:
    \begin{itemize}
      \item
        Granularity:\\
        People will make reservations during times that are logical to them, 
        hence the smallest interval can be $1$ minute, 
        $5$ minutes, $30$ minutes, $1$ hour and so on, depending on the application.
      \item
        Starting points:\\
        All interval sizes in the tree should be whole minute intervals. For 
        instance, if an interval of $15$ minutes should be divided, the divisor $2$
        is not a proper choice since the children will then have an interval size
        of $7.5$ minutes and those children will in turn be divided to $3.75$ minutes, which both will 
        rarely occur. The divisor $3$ is in this example a more suitable divisor, which 
        will make the $15$ minutes interval divided in $3$ intervals of size $5$ minutes each.
      \item
        Size of intervals:\\
        If an interval is estimated to be more likely than another, use the more likely choice.
        For instance, if a $24$ hour day is going to be divided, the divisors $2$ and $3$ seem like 
        good choices. If we estimate that it is more likely that reservations of $8$
        hours will occur rather than $12$ hours, due to the fact that a working day of humans 
        is $8$ hours, the divisor $3$ should be chosen. This choice is however only an improvement
        if it makes the starting time and the ending time of the interval the
        same as the working hours.
    \end{itemize}
    An example of a tree divided with respect to humans are shown in \figref{figures1}.
    
  \subsubsection{Early rejection}
  \sectlabel{section:Early rejection}
  
    The sooner we can detect that a reservation will not fit, due to the fact that 
    there is not enough bandwidth left to reserve, the better. In the description 
    of the $\mathtt{Insert}$ function we assumed that the check is done independently of the 
    insertion. In this section we give some hints how to incorporate the check into the $\mathtt{Insert}$ function

  \subsubsection*{Inclusion of check within insertion}

    In order not to reserve more bandwidth than can be delivered the $\mathtt{Insert}$ function 
    has to check for over-reservations. One way to do this is when the $\mathtt{Insert}$ 
    function discovers that there will be an over-reservation, the recursion stops, 
    and undo of the insertions in the nodes so far is performed.
    Another way is that the 
    recursion stops, marks the current node, and then a clean up function is called with the same
    arguments as the $\mathtt{Insert}$ function before. The clean up function can be 
    implemented as a $\mathtt{Delete}$ function that 
    only deletes the values up to the marked node and then stops. The second approach gives 
    simpler code, fewer boolean checks, and therefore is also somewhat faster.
    

  \subsubsection*{Order of traversal}

    The probability that a new reservation will not fit within the reservable bandwidth 
    is somehow bigger at larger intervals. Hence during the recursion, if an interval has to 
    be divided into smaller parts,
    the recursion first follows the largest interval of the edge parts of 
    the interval, that is $\max\left( I_{Q_1}, I_{Q_{rightmost}}\right) $ 
    (see \figref{figures4}).

\subsubsection{Locality}

  Since our tree is spanning a large time interval, we can assume that there will be some 
  form of locality in the reservations and queries made.
  The locality gives additional improvements due to the fact that parts of generated data 
  used during previous operations is already in the cache (cf. cache-aware data structures \cite{arge-ioefficient,arge-cacheoblivious,arge-cacheoblivious2})
  For our data structure we have developed a method of start the traversal of the tree 
  inside the tree and not always from the top. In order to do that we need to find the lowest
  node that spans both the last interval, $I_{Last}$ and the current 
  interval, $I_{Current}$, that is the lowest node that spans the interval 
  $I_{Merged}=\left[ \min\left( S_{Last},S_{Current} \right) , \max\left( E_{Last},E_{Current} \right) \right] $.
  This can be achieved by using the set $X$ and a table of $\Abs{M_l}$ 
  (\rEqu[eq.]{Intervalsize}). The operations on our data structure collect information from the
  top node and down and we need to maintain this information even when starting from the inside of the tree. 
  This 
  is done by using a stack in which all accumulated $nv$'s are stored during the recursion down to the 
  uppermost node in which $I_{Last}$ was divided. When, for instance, querying the interval $I_{Current}$,
  we do a binary search among the $\left( L=O\left( \log n\right) \right) $ levels in the stack 
  to find the level, $l$, that has a node, $N$, spaning $I_{Merged}$. Then the recursion starts 
  in $N$ with the start-$nv$ from level $l$ in the stack. The running time of the search for 
  $N$ is $O(\log \log n)$.
  
\subsubsection*{Similar results}

    There is a
    resemblance between the searching in this data structure and a 
    repeated search in a B-tree. Guiba et al. (\cite{GuibasMPR77})
    have proposed an algorithm to traverse a B-tree without always
    having to start from the top node. They are using a lot of fingers
    (pointers) between nodes in the tree and a special search pattern. They
    use the search pattern to find the lowest common ancestor.
    If the node to find is to the right of the last node the pattern will be:
    \begin{itemize}
      \item
        ``go to right neighbour'', ``go to father'' and so on.
    \end{itemize}
    If the next node is to the left of the last node the pattern is:
    \begin{itemize}
      \item
        ``go to left neighbour'', ``go to father'' and so on. 
    \end{itemize}
    Since in our operations the node value needs to be accumulated from the root 
    and down on the search path, 
    Guiba et. al.'s proposal can not be directly used in our case. Further, the idea to perform
    a binary search on the complete path from the root to a given node is not new and 
    was already employed in \cite{GonnetMunro1986}.

\subsubsection{Time goes}

  In our modified segment tree the nodes are not moved into new positions and
  new nodes are not inserted, neither are nodes in the tree deleted. Hence, the
  time frame of the tree is static. But as time moves ahead towards the future,
  more reservations will be inserted more and more to the right parts of the tree.
  Eventually the intervals of the reservations will be so far off into the future that 
  the interval will miss the data structure. 
  A solution is to make the data structure wrap the spanning interval. Let the 
  spanning interval of the data structure be twice as large as is believed to be 
  needed. Then when the current time passes the first half 
  of the entire interval, the interval is wrapped so that the first half becomes the third, and so on.
    
\section{Conclusions}\sectlabel{Discrete_Conclusions}
  In this paper we presented a discrete data structure for 
  reservations of limited resources over a time-span. The data structure is made
  with bandwidth reservations on the Internet in mind, but it is quite generic. It can be applied whenever a 
  limited resource shall be reserved. The data structure proposed has time complexity 
  independent of the number of reservations
  made and the size of the interval. We defined three operations on the data structure;
  $\mathtt{Insert}$ to insert a reservation, $\mathtt{Delete}$ to delete a reservation, 
  and $\mathtt{MaxReserved}$ to query how much bandwidth is used during a query 
  interval. 
  The worst case time complexity of all operations is $O(\log n)$. In the second part of 
  the paper we simplified the data structure by making it implicit.
  We conclude the paper with a number of suggestions how to improve the solution.
  

\bibliography{advanced}
\bibliographystyle{plain}
\end{document}